\documentstyle[aps,preprint,prc,12pt]{revtex}

\begin{document}

\title{
Searching for the Nuclear Liquid-Gas Phase Transition in Au + Au Collisions at
35 MeV/nucleon
}

\author{
M. Belkacem$^{1,2}$, P.F. Mastinu$^{1,3}$, V. Latora$^2$, A. Bonasera$^2$,
M. D'Agostino$^1$, M. Bruno$^1$, J. D. Dinius$^4$, M. L. Fiandri$^1$, 
F. Gramegna$^5$, D. O. Handzy$^4$, W. C. Hsi$^4$, M. Huang$^4$, M. A. Lisa$^4$,
G. V. Margagliotti$^6$, P. M. Milazzo$^6$, C. P. Montoya$^4$, 
G. F. Peaslee$^4$, R. Rui$^6$, C. Schwarz$^4$, G. Vannini$^6$ 
and C. Williams$^4$
}

\address{
$^{1}$ Dipartimento di Fisica and INFN, Bologna, Italy \\
$^{2}$ INFN, laboratorio Nazionale del Sud, Catania, Italy \\
$^{3}$ Dipartimento di Fisica, Padova, Italy \\
$^{4}$ NSCL, Michigan State University, USA \\
$^{5}$ INFN, Laboratori Nazionali di Legnaro, Italy \\
$^{6}$ Dipartimento di Fisica and INFN, Trieste, Italy 
}

%\date{ \today }
\maketitle

%\newpage

\begin{abstract}

Within the framework of Classical Molecular Dynamics, we study the collision
Au + Au at an incident energy of 35 MeV/nucleon. It is found that
the system shows a critical behaviour at peripheral impact parameters,
revealed through the analysis of conditional moments of charge distributions,
Campi Scatter Plot, and the occurrence of large fluctuations
in the region of the Campi plot where this critical behaviour is expected.
When applying the experimental filters of the MULTICS-MINIBALL apparatus, it is
found that criticality signals can be hidden due to the inefficiency
of the experimental apparatus. The signals are then recovered by identifying
semi-peripheral and 
peripheral collisions looking to the velocity distribution of the
largest fragment, then by selecting the most complete events.

\end{abstract}

{
\vskip 2\baselineskip
}

% body of the paper
\newpage

%\narrowtext
%\begin{multicols}{2}

\section{Introduction}

From several years, the idea that nuclear systems may show up some evidence
for the occurrence of a critical behaviour related to a liquid-gas phase 
transition has stimulated lots of investigations both from theoretical and
experimental sides \cite{curt,berts,good,jaq,palm,daniel,lamb,jaq1}. 
This idea has initiated more than ten years ago with the
observation by the Purdue-Fermilab group of asymptotic fragment charge 
distributions
exhibiting a power law \cite{purd}. 
Such a power law, as described by the Fisher's droplet 
model, is expected for cluster formation near the critical point of a 
liquid-gas phase transition \cite{fisher}. 
This interest increased recently with the attempt by the EOS Collaboration to
extract critical exponents of fragmenting nuclear systems produced in the
collision of 1 GeV/nucleon Au nuclei with a carbon target \cite{eos}, and
with the extraction by the ALADIN Collaboration of a caloric curve resulting
from the fragmentation of the quasiprojectile formed in the collision
Au + Au at 600 MeV/nucleon exhibiting a behaviour expected for a liquid-gas
phase transition \cite{aladin}.

In the present report, we study within the framework of Classical Molecular 
Dynamics (CMD) model, the reaction $^{197}$Au + $^{197}$Au at an 
incident energy of 35 MeV/nucleon, then 
analyse the results in terms of critical behaviour by studying fragment charge
distributions, their moments, and the occurrence of large
fluctuations in terms of intermittency analysis, and as shown by the 
fluctuations of the size of the largest fragment.
Our aim for this analysis is three fold. First, a critical behaviour has been
observed in the fragmentation of Au nuclei in the previously mentioned 
experiments at high beam energies (600 and 1000 MeV/nucleon), and we are 
interested to see if such a behaviour can still be observed at an energy 
notably
lower. Second, we want to identify the critical events (if any) and study the
effects of the efficiency of experimental apparatus (by applying for example 
the
experimental filters of the MULTICS-MINIBALL apparatus \cite{iori,desouza}). 
Doing this we can see
whether experimental inefficiencies can completely wash out 
the signals of criticality or it is still 
possible to recover these signals from the 
filtered results. Finally, we aim to apply the same procedure to the
experimental data obtained by the MULTICS-MINIBALL collaboration for the same 
reaction \cite{central}.

This paper is organized as follows. We give in section II a brief description
of the CMD model used for this study. A more complete description can be found
in Refs. \cite{pan90,belk}. Section III contains the analysis of the moments
of charge distributions, the Campi scatter plot and the analysis of the scaled
factorial moments in terms of the intermittency signal. Section IV is devoted
to the study of the effects of experimental inefficiencies, applying to the CMD
results the filters of the MULTICS-MINIBALL apparatus, and how it is possible 
to recover the signals of criticality selecting well detected events. 
Finally, we give in section V our summary and conclusions.

\section{Brief description of the CMD model}

In the CMD model, we assume that each nucleus is made up of 197 nucleons (79
protons + 118 neutrons) that move under the influence of a two-body potential
$V$ consisting of two different interactions \cite{pan90}:
\begin{eqnarray}
V_{nn}(r) &=& V_{pp}(r) = V_{0}\left[exp(-\mu_{0}r)/{r} -
exp(-\mu_{0}r_{c})/{r_c}\right]
\nonumber \\
V_{np}(r) &=& V_{r}\left[exp(-\mu_{r}r)/{r} - exp(-\mu_{r}r_{c})/{r_c}\right]
\nonumber \\
          & & \mbox{}- V_{a}\left[exp(-\mu_{a}r)/{r} -
exp(-\mu_{a}r_{a})/{r_a}\right]
\label{eq1}
\end{eqnarray}
${r_c}=5.4~ fm $ is a cutoff radius.
The first interaction, for identical 
nucleons, is purely repulsive so no bound state of identical nucleons can exist
(to simulate in some sense the Pauli principle), and the second, for 
proton-neutron
interaction, is attractive at large distances and repulsive at small ones
\cite{pan90}. 
The various parameters entering Eq.(1) are defined with their respective
values in Ref. \cite{pan90}.
This potential gives an Equation of State (EOS) of 
classical matter having about 250 MeV of compressibility
(set M in Ref. \cite{pan90}), and which
strikingly resembles that of nuclear matter (i.e. equilibrium
density $ \rho_{0} = 0.16$ fm$^{-3}$ and energy
$E( \rho_{0}) = -16$ MeV/nucleon).
Furthermore, in Refs. \cite{pan90,delzoppo},
it is shown that many experimental data on
heavy-ion collisions are reasonably explained by this classical model.
Of course this is not accidental but it is due
to the accurate choice of the parameters
of the two-body potentials \cite{pan90}.
The classical Hamilton's equations of motion are solved using the Taylor
method at the order O[$(\delta t)^3$] where $\delta t$ is the integration
time step \cite{computer}.  Energy and momentum are well conserved. Both nuclei
are initialized in their ground state by using the frictional cooling
method \cite{fcm}, then they are boosted towards each other with the CM kinetic
energy.
In the present calculations, the Coulomb interaction is 
explicitly taken into account. Note that even though this model is completely
classical, it solves exactly (within the numerical error bars) the classical
many-body problem, taking into account all order correlations.

\section{Results}

Calculations for the reaction Au + Au at 35 MeV/nucleon are carried out for 
several impact parameters,
from 1 to 13 $fm$ by steps of 1 $fm$. One intuitively imagines the following 
scenario for this reaction. For central collisions, since the incident energy
is rather high as compared to the Coulomb barrier, the two heavy nuclei will 
come in contact for a short time. The total charge of the intermediate system
is very high and it will quickly explode due mainly to the high Coulomb 
repulsion \cite{central}.
For increasing impact parameter, two or may be three excited primary fragments
might be formed. By tuning the impact
parameter, we might hope to obtain some primary sources which have a
combination
of excitation energy, Coulomb charge, and angular momentum sufficient to bring
the system into the instability region (if any). The possibility that such a
scenario might apply to heavy-ion collisions has been shown in microscopic
calculations \cite{belk,gross}.
In particular, it has been shown that the "critical" excitation
energy decreases when the system is either charged and/or rotating
\cite{belk,gross,bondorf,bonasera}. Thus a
combination of all these ingredients might give the desired result. 
Following this scenario, one would expect to see a critical behaviour (if any)
for peripheral collisions.

In Fig. 1, we have plotted the dynamical evolution in the $x-z$ plane
for this reaction for four different times (after the two nuclei came in 
contact), and four different impact parameters;
$\hat{b}=0.15$ (first line panels), $\hat{b}=0.38$ (second line panels), 
$\hat{b}=0.62$ (third line panels) and
$\hat{b}=0.85$ (fourth line panels). 
For central and semi-central collisions (first and second rows)
the two nuclei come in contact with each other and form a unique deformed 
source (the source is less deformed for more central collisions) which decays 
through light particles and fragments emission \cite{central}.
For semi-peripheral and 
peripheral collisions (third and fourth lines panels), 
one sees clearly the formation of two 
big sources (the quasitarget and quasiprojectile) with the formation between 
them of a third smaller source in the neck region. The size of this "neck"
is smaller for more peripheral collisions and it completely disappears for the 
most peripheral ones.

One of the most powerful methods used to characterize the critical behaviour
of a system undergoing a multifragmentation is the method of conditional 
moments introduced by Campi \cite{campi}. 
The moments of asymptotic cluster charge distributions are 
defined as:
\begin{equation}
m^{(j)}_{k} = \sum_{Z} Z^{k} n^{(j)}(Z)/Z_{tot}
\label{mm}
\end{equation}
where $n^{(j)}(Z)$ is the multiplicity of clusters of charge $Z$ 
in the event $j$, $Z_{tot}=158$,
and the summation is over all the fragments in the event {\it except the
heaviest one}, which corresponds to the bulk liquid in an infinite system. 
If the system keeps some trace of the phase transition for some
particular events, the moments $m_{k}$ should
show up some strong correlations between them \cite{campi}. In particular, the 
second 
moment $m_{2}$, which in macroscopic thermal systems is proportional to the 
isothermal compressibility, diverges at the critical temperature 
\cite{eos,balescu,fino}. 
Of course
in finite systems, the moments $m_{k}$ remain finite due to finite size
effects. In the upper part of Fig. 2, we have plotted versus the reduced impact 
parameter $\hat{b}$
the second moment $m_{2}$, calculated taking out the two largest
fragments instead of only the largest one because, if one expects the critical
behaviour at peripheral impact parameters and as the system is symmetric, 
one should subtract both bulk 
fragments coming from the quasitarget and the quasiprojectile. As 
expected, the second moment $m_{2}$ shows a peak for an impact parameter
$\hat{b} \approx 0.8$. If one
does not take off the second largest fragment (lower part of Fig. 2), we 
observe a continuous rise of $m_{2}$ and the peak disappears  
because we are summing with small fragments, a very big one (bulk) 
to the square (or power $k$ for the highest moments $m_{k}$). 
In Fig. 3, we have plotted the same quantity $m_{2}$ versus the multiplicity 
of charged particles $N_{c}$ (with $Z \ge 1$), calculated without the two
largest fragments (upper part) and only without the largest one (lower part).
The second moment $m_{2}$ shows also a peak versus $N_{c}$ for a multiplicity
around 20-25, 
and this peak disappears when taking into account the second largest 
fragment. In the following, the analysis of the non-filtered results is done 
taking off the two largest fragments.

Another quantity proposed by Campi to give more insight into the
critical behaviour is the relative variance $\gamma_{2}$ defined as 
\cite{campi}:
\begin{equation}
\gamma_{2} = \frac{m_{2}m_{0}}{m_{1}^{2}}
\label{gam2}
\end{equation}
It was shown by Campi that this quantity presents a peak around the
critical point which means that the fluctuations in the fragment size
distributions are the largest near the critical point \cite{campi}.
In Fig. 4, we have plotted the relative variance $\gamma_{2}$
versus the reduced impact parameter $\hat{b}$ (upper part), and versus the 
charged particle multiplicity $N_{c}$ (lower part). One clearly notes that 
the relative variance $\gamma_{2}$ shows
a peak in both plots, for a reduced impact parameter $\hat{b}$ around 0.8, and 
for $N_{c} \approx 20-25$.

Moreover, we have considered another variable which is the normalized variance 
of the charge of the maximum fragment $\sigma_{NV}$. As charge distributions
are expected to show the maximum fluctuations around the critical point
\cite{stanley}, this quantity is expected to present some maximum at the 
critical point \cite{campi,dorso}
This normalized variance is defined as
\begin{equation}
\sigma_{NV} = \frac{\sigma^{2}_{Z_{max}}}{<Z_{max}>}
\label{nv}
\end{equation}
where 
\begin{equation}
\sigma^{2}_{Z_{max}} = <Z_{max}^{2}> - <Z_{max}>^{2}
\end{equation}
The brackets $<~^.~>$ indicate an ensemble averaging.
We have plotted in Fig. 5 the normalized variance $\sigma_{NV}$ versus 
$\hat{b}$ (upper part), and versus $N_{c}$ (lower part). In this case also, 
we observe a peak for this quantity in both plots at
almost the same values of $\hat{b}$ and $N_{c}$.
This means that the fluctuations in the 
charge of the maximum fragment (thus in charge distributions) 
are the largest around these values of the 
impact parameter and charged particle multiplicity.

The upper part of Fig. 6 
shows a scatter plot of $ln(Z^{j}_{max})$ versus $ln(m^{j}_{2})$ for 
each event $j$,
commonly known as Campi scatter plot. It was shown that if the system keeps 
some trace of the phase transition, the correlation between these two 
quantities exhibits two characteristic branches, an upper branch with an
average negative slope corresponding to undercritical events, and a lower 
branch with a positive slope that corresponds to overcritical events, and
the two branches meet close to the critical point of the phase transition
\cite{belk,campi,gross1}.
The results of Fig. 6 show two branches corresponding to undercritical and
over critical events, similar to the predicted ones. Note that the upper 
branch is made mainly by events having an impact parameter $\hat{b} > 0.85$,
while the lower branch is made by events having $\hat{b} < 0.77$. The region
where the two branches meet is made by events having 
$0.77 \le \hat{b} \le 0.85$. 
In the following, we will show that the central region where the two branches
meet and where the critical behaviour is expected, is characterized by the 
occurrence of large fluctuations, revealed through an 
intermittency analysis \cite{bialas}. 
In the lower part of Fig. 6, we have plotted the 
logarithm of the scaled factorial moments (SFM) defined as \cite{plocia}
\begin{equation}
F_i(\delta s)={{\sum _{k=1}^{Z_{tot}/ \delta s}<{n_k}\cdot ({n_k}-1)\cdot ...
\cdot({n_k}-i+1)>}
\over {\sum _{k=1}^{Z_{tot}/ \delta s}<n_k>^i}}
\label{SFM}
\end{equation}
$i=2,...,5$ versus the logarithm of the bin size $\delta s$. In the above
definition of the SFM, $i$ is the
order of the moment. The total interval $1-Z_{tot}$ ($Z_{tot} = 158$)
is divided in $M={Z_{tot}/ \delta s}$ bins of size $\delta s$,
${n_k}$ is the number of particles in the $k$-th bin for an event,
and the brackets $<~^.~>$ denote the average over many events. An intermittent 
pattern of fluctuations is characterized by a linear rise of the logarithm of 
the SFM's versus $-ln(\delta s)$ (i.e. $F_{i} \propto
\delta s^{-\lambda_{i}}$)
which corresponds to the existence of large  fluctuations which
have self-similarity over the whole range of scales considered
\cite{gross1,bialas,plocia}.
Even though this quantity is ill defined for fragment distributions
\cite{campi1,phair}, it has been shown in several
theoretical studies that critical events give a power law for the SFM versus
the bin size \cite{belk,delzoppo,plocia,hwa,kubo}. In the figure, the 
logarithm of 
the SFM's
exhibits a linear rise versus the logarithm of the bin size indicating a strong
intermittency signal in the region of the Campi plot where the critical 
behaviour is expected. To understand whether these large fluctuations are due 
to a simple event mixing by considering different impact parameters inside Cut
2, we fixed the impact parameter to say $\hat{b} = 0.85$. The resulting SFM 
are shown in Fig. 7. One notes that the signal is still there even much 
weaker than previously (the absolute values of the SFM are smaller). This 
allows us to conclude that the intermittency signal is not due to the mixing 
of events and this mixing only increases the absolute values of the SFM.

At the end of this section, we would like to say few words about the mixing of 
different sources in the calculations of the previous quantities. First of all,
we note that it is not evident to separate the different sources which might 
be formed after the first stages of the collision when they are still 
overlapping (we mean by overlap distances smaller than the range of the two-body
interaction used, i.e. $r_{c}=5.4$ fm), as one can see from Fig. 1. Thus it is
not obvious to distinguish which fragments come from the different sources, 
even for a simple dynamical model like CMD. For the calculations of the second 
moment $m_{2}$ for instance, one should consider only one source (that entering
the critical region). For central collisions, only one source is formed and 
$m_{2}$ is calculated according to Eq.(\ref{mm}) with $Z_{tot} \approx 158$. 
For peripheral impact parameters, one should calculate the second moment only
from one source (the PLF or TLF assuming two sources), and in this case 
$Z_{tot}$ should be around 79 (158/2) in Eq.(\ref{mm}). As we are dealing with
a symmetric reaction, we can say that both the PLF and the TLF enter separately
the critical region. So calculating $m_{2}$ using Eq.(\ref{mm}) with 
$Z_{tot} \approx 158$ is equivalent calculating it by suming on the fragments
coming from only one source and dividing by $Z_{tot} \approx 79$, which gives 
the same results as those of Fig. 2. This discussion holds for all the moments
$m_{k}$, thus for the reduced variance $\gamma_{2}$. For the normalized 
variance of the charge of the largest fragment, one should be careful to 
consider the largest fragment coming from only one source (this was not done 
for the previous calculations of $\sigma_{NV}$). For central collisions, we 
have only one source, and the results do not change. For peripheral collisions,
by considering only the largest fragment with a positive velocity in the centre
of mass, the obtained peak in $\sigma_{NV}$) is higher than that obtained
previously (3.8 instead of 2.4 of Fig. 4). This result is in some sense obvious
because we were previously smoothing the fluctuations of the largest fragment
on both sources (PLF and TLF). For the Campi plot, we have plotted the 
logarithm of the size of the largest fragment versus the logarithm of $m_{2}$
both calculated for the fragments emitted in the forward direction (with
$v_{CM} \ge 0$ to select roughly the PLF source). The obtained results are 
very similar to those reported on Fig. 6 and making a gate on the central region
of the plot, we obtained almost the same SFM with the same absolute values as
those reported on the lower part of Fig. 6.

In this section, we have seen that the analysis of the reaction Au + Au at 
35 MeV/nucleon shows a signal of critical behaviour in peripheral collisions.
This behaviour is revealed through the analysis of the second moment of 
charge distributions, the reduced variance, the large 
fluctuations of the size of
the largest fragment, the characteristic shape of the Campi scatter plot and 
the occurrence of large fluctuations in the region of the Campi 
plot where the critical behaviour is expected.

\section{ Effects  of Experimental Inefficiency}

As indicated in the introduction, one of the aims of this study is to apply
the same procedure of critical behaviour identification to the experimental
data obtained by the MULTICS-MINIBALL Collaboration for the same reaction, Au 
+ Au at 35 MeV/nucleon. To do so, we have filtered our results using the 
angular acceptance and energy thresholds of the MULTICS-MINIBALL apparatus. 

First of all, we have checked that at least for semi-peripheral and peripheral
collisions, the efficiency of the apparatus automatically eliminates the
largest fragment coming from the target-like, so we calculate the moments
of charge distributions $m_k$ (Eq.(\ref{mm})) by subtracting only the largest
fragment (and not the two largest ones as for the unfiltered results).
The upper part of Fig. 8 shows the second moment $m_{2}$ versus 
$\hat{b}$. The second moment $m_{2}$ does no more show the peak observed for
unfiltered results around $\hat{b}= 0.8$, even though one notes some remaining
of that peak. We note also the appearance of a bump for more central 
collisions, around $\hat{b}=0.38$. The 
situation is worst for the plot of $m_{2}$ versus $N_{c}$ in the lower panel
of the figure
where one observes
only a quasi-linear rise. The reduced variance $\gamma_{2}$ drawn in Fig. 9, 
shows a bit different behaviour. One still observes a smooth bump at $\hat{b}
=0.8$ but $\gamma_{2}$ is almost constant for $\hat{b} < 0.8$ and not rising
as it is 
the case for unfiltered results. The same is for the plot of $\gamma_{2}$ 
versus $N_{c}$. The normalized variance of the size of the largest fragment
represented in Fig. 10, still shows a peak but a little shifted versus higher
impact parameters (upper part of the figure, see Fig. 4) and lower charged
particle multiplicity (lower part). More drastic is the change in the shape 
of the Campi scatter plot shown in Fig. 11. This plot does no more show any
particular shape characteristic to the occurrence of a critical behaviour
(observed in the unfiltered results) and one is no more able to identify the
upper and lower branches neither the meeting zone.

The effects of apparatus inefficiencies can thus be more or less drastic 
depending on the variable we are looking at. To 
recover the signals of criticality, we adopted the following procedure: 

{\it i)} As the critical behaviour was observed at peripheral impact 
parameters, we identify semi-peripheral and
peripheral collisions, eliminating more central ones, by selecting those events
in which the velocity of the largest fragment along the beam axis is larger or 
equal to 75$\%$ of the beam velocity, which means that we are selecting those
events in which there is a remnant of the projectile flying with the velocity
of the quasiprojectile. Doing this, we hope to select only those reactions 
where 
two or three primary sources are formed (semi-peripheral and peripheral 
reactions) and eliminate the reactions where only one source is formed at 
mid-rapidity (central collisions); 

{\it ii)} we select the most complete events imposing that the 
total detected charge is larger than 70 ($Z_{tot} \ge 70$). 

\noindent
Moreover, we have checked that 
condition {\it i)} does not automatically eliminate all central collisions and 
to do so one needs to impose a maximum limit to the total detected charge, say
$Z_{tot} \le 90-95$. We note also that changing condition {\it i)} from 75$\%$
to 85$\%$ of the beam velocity for example does not change 
significantly the results, and only decreases the statistics. 

In Figs. 12, 13 and 14, we have plotted the second moment $m_{2}$, the reduced
variance $\gamma_{2}$ and the normalized variance $\sigma_{NV}$ versus the 
reduced impact parameter $\hat{b}$ (upper part of the figures) and versus
charged particle multiplicity $N_{c}$ (lower part). One sees that the signals
observed for non-filtered results are recovered at the same impact parameter.
One notes also that this selection eliminates central collisions with
$\hat{b} \le 0.38$. 

Figure 15 displays the Campi scatter plot for the filtered events with the 
selection on the velocity of the largest fragment and the total detected
charge. We see that one recovers the characteristic shape of the Campi plot, 
in that it shows the upper branch 
with a negative slope and the lower branch with a positive slope, already 
observed in the unfiltered results. 
To better clarify the characteristics of these two branches and of the 
meeting zone, we have made three cuts in this plot selecting the
upper branch (Cut 1), the lower branch (Cut 3) and the
central region (Cut 2), and analysed the events falling in each of the 
three cuts. 
The upper part of Fig. 16 shows the impact parameter distributions
of the events falling in the three cuts of the Campi plot. One sees that the 
three cuts select different regions of the impact parameter distribution; 
Cut 1 (left panel) 
selects the 
most peripheral collisions with a distribution peaked at $\hat{b}=0.92$, Cut
2 (central panel) selects peripheral impact parameters with a distribution 
going from 
$\hat{b}=0.65$ to 0.95 while Cut 3 (right panel) selects more central 
collisions. In the
lower part of the same figure, we have plotted the charged particles 
multiplicity distributions for the three cuts. Cut 1 shows a multiplicity
distribution from 2 to 10 while Cut 3 shows a distribution at higher 
multiplicities from 30 to 45. The situation is different for Cut 2. 
The multiplicity distribution covers a wider range of $N_{c}$ values from 2 to 
30. Note that this large multiplicity distribution is not due, as one can 
think, to a large impact parameter mixing (see upper part of the figure), but
can be due to the occurrence of large fluctuations (as we will see) as expected
near the critical point. 

Figure 17 displays in the upper part the fragment
charge distributions obtained in the
three cuts \cite{note}
with, in the lower part, the corresponding scaled factorial moments 
calculated according to Eq.(\ref{SFM}) \cite{note}.
Cut 1 (left part of the figure) corresponds
to undercritical events and hence one obtains a charge distribution with a
"U" shape characteristic to evaporation events, while for Cut 3 (right
part) one observes a rapidly decreasing charge distribution with an 
exponential shape characteristic to highly excited systems going to 
vaporization. For
Cut 2 (central part), we obtain a fragment charge distribution exhibiting
a power law $Z^{-\tau}$, with $\tau \approx 2.2$, which is
expected, according to
the droplet model of Fisher, for fragment formation near the critical point
indicating a liquid-gas phase transition, and consistent with the scaling
laws of critical exponents \cite{fisher}.
In the lower part of the figure, for region 1 corresponding to evaporation
events, the logarithms of the scaled factorial moments $ln(F_i)$
are always flat and independent on $-ln(\delta s)$ and there is 
no intermittency signal. For Cut 2 the situation is different. 
The logarithms of the SFM's are positive and almost linearly increasing versus 
$-ln(\delta s)$ and a strong intermittency signal is observed (note the 
absolute values of $ln(F_i)$). Cut 3 gives negative logarithms of the SFM's
and we have also in this case no intermittency signal. Note that this
behaviour of the scaled factorial moments is exactly the same as that observed
in percolation and molecular dynamics models for undercritical, critical and 
overcritical events, respectively \cite{belk,baldo}.

\section{Conclusions and Outlooks}

In conclusion, we have studied the reaction Au + Au at an incident energy 
of 35 MeV/nucleon within the framework of Classical Molecular Dynamics. 
The results show evidence for the occurrence of a critical behaviour revealed
through the shape of the second moment of charge distributions, the reduced
variance, the normalized variance of the size of the largest fragment, the
particular shape of the Campi scatter plot and through the presence of large 
 fluctuations as indicated by the intermittency analysis in the 
region of the Campi plot where the critical behaviour is expected. We have 
also seen that when our results are filtered using the geometrical acceptance
and energy thresholds of the MULTICS-MINIBALL apparatus, experimental 
inefficiencies can hide more or less the signals of criticality. Moreover, we 
have shown that these criticality signals can be recovered by identifying the
most complete semi-peripheral and peripheral events selecting those events in 
which the largest fragment has a velocity along the beam axis larger or equal
to 75$\%$ of the beam velocity and for which the total detected charge is
$70 \le Z_{tot} \le 90$. 

We would like to note at the end that the same procedure for characterizing
the critical behaviour has been successfully applied to the experimental data
obtained by the MULTICS-MINIBALL Collaboration for the same reaction Au + Au 
at 35 MeV/nucleon, and that a critical behaviour has been identified 
\cite{prl}. As an example, we show in Fig. 18 the experimental Campi scatter
plot \cite{prl} obtained making more or less the same event selection as for 
the CMD results. Note the strong similarity with the theoretical Campi plot
shown in Fig. 15. Moreover, we show in Fig. 19 the experimental scaled 
factorial moments \cite{prl} obtained in the three cuts made on the Campi plot
of Fig. 18. Once again note the similarity of these results with those of the 
CMD results.
The authors of the previous reference 
have also extracted the other quantities discussed in
this paper (variance of the charge of the largest fragment, etc..) from
the experimental data \cite{bormio96}. These quantities behave very similarly
to what is discussed here for the CMD case thus strengthening
our findings.  A very similar behaviour to the one discussed
here has also recently been observed in Xe + Sn collisions at 55 MeV/nucleon
measured with the detector INDRA again for peripheral collisions 
\cite{benlliure}.
Work now is
in progress to characterize the fragmenting sources leading to the critical 
behaviour and maybe to extract the critical exponents.

%\newpage
{
\vskip 0.7 cm
\centerline{\bf ACKNOWLEDGMENTS}
}

One of us (M. Belkacem) thanks the Physics Department of the 
University of Trieste for financial support and the Physics Department of the
University of Bologna, where part of this work has been done, for warm 
hospitality and financial support.

%\end{multicols}

\newpage

\begin{figure}
\label{f1}
\noindent
\caption{Dynamical evolution. The r-space distribution is projected on the
$x-z$ plane.}
\end{figure}
 
\begin{figure}
\label{f2}
\noindent
\caption{The second moment of charge distributions $m_{2}$ versus the reduced 
impact parameter $\hat{b}$. The upper panel: without the two largest fragments, 
the lower panel: without only the largest fragment.}
\end{figure}
 
\begin{figure}
\label{f3}
\noindent
\caption{The second moment of charge distributions $m_{2}$ versus charged 
particle multiplicity $N_{c}$. The upper panel: without the two largest 
fragments, the lower panel: without only the largest fragment.}
\end{figure}
 
\begin{figure}
\label{f4}
\noindent
\caption{The reduced variance $\gamma_{2}$ versus the reduced impact parameter
$\hat{b}$ (upper panel) and versus charged particle multiplicity $N_{c}$ (lower
panel). The calculations are done without the two largest fragments.}
\end{figure}
 
\begin{figure}
\label{f5}
\noindent
\caption{The normalized variance of the size of the largest fragment 
$\sigma_{NV}$ versus the reduced impact parameter $\hat{b}$ (upper panel) and 
versus charged particle multiplicity $N_{c}$ (lower panel).}
\end{figure}
 
\begin{figure}
\label{f6}
\noindent
\caption{Upper panel: Campi scatter plot. The logarithm of the size of the 
largest fragment $ln(Z_{max})$ is plotted versus the logarithm of the second 
moment $ln(m_{2})$. Lower panel: The logarithm of the scaled factorial moments
$ln(F_i)$ is plotted versus the logarithm of the bin size $-ln(\delta s)$ for 
the events falling within the cut drawn in the Campi scatter plot, upper panel.
Solid circles represent the SFM of order $i=2$, open circles $i=3$, open 
squares $i=4$ and open triangles $i=5$.}
\end{figure}

\begin{figure}
\label{f7}
\noindent
\caption{The logarithm of the scaled factorial moments
$ln(F_i)$ is plotted versus the logarithm of the bin size $-ln(\delta s)$ for 
the events with $\hat{b} = 0.84$.
Solid circles represent the SFM of order $i=2$, open circles $i=3$, open 
squares $i=4$ and open triangles $i=5$.}
\end{figure}
 
\begin{figure}
\label{f8}
\noindent
\caption{Filtered CMD results. The second moment of charge distributions 
$m_{2}$ versus the reduced impact parameter $\hat{b}$ (upper panel) and
versus charged particle multiplicity $N_{c}$ (lower panel).}
\end{figure}
 
\begin{figure}
\label{f9}
\noindent
\caption{Filtered CMD results. The reduced variance $\gamma_{2}$ versus the 
reduced impact parameter $\hat{b}$ (upper panel) and versus charged particle 
multiplicity $N_{c}$ (lower panel).}
\end{figure}
 
\begin{figure}
\label{f10}
\noindent
\caption{Filtered CMD results. The normalized variance of the size of the 
largest fragment $\sigma_{NV}$ versus the reduced impact parameter $\hat{b}$ 
(upper panel) and versus charged particle multiplicity $N_{c}$ (lower panel).}
\end{figure}
 
\begin{figure}
\label{f11}
\noindent
\caption{Filtered CMD results. Campi scatter plot. The logarithm of the size 
of the largest fragment $ln(Z_{max})$ is plotted versus the logarithm of the 
second moment $ln(m_{2})$.}
\end{figure}
 
\begin{figure}
\label{f12}
\noindent
\caption{Filtered CMD results with selection of events. The second moment of 
charge distributions
$m_{2}$ versus the reduced impact parameter $\hat{b}$ (upper panel) and
versus charged particle multiplicity $N_{c}$ (lower panel).}
\end{figure}
 
\begin{figure}
\label{f13}
\noindent
\caption{Filtered CMD results with selection of events. The reduced variance 
$\gamma_{2}$ versus the
reduced impact parameter $\hat{b}$ (upper panel) and versus charged particle
multiplicity $N_{c}$ (lower panel).}
\end{figure}
 
\begin{figure}
\label{f14}
\noindent
\caption{Filtered CMD results with selection of events. The normalized 
variance of the size of the
largest fragment $\sigma_{NV}$ versus the reduced impact parameter $\hat{b}$
(upper panel) and versus charged particle multiplicity $N_{c}$ (lower panel).}
\end{figure}
 
\begin{figure}
\label{f15}
\noindent
\caption{Filtered CMD results with selection of events. Campi scatter plot. 
The logarithm of the size
of the largest fragment $ln(Z_{max})$ is plotted versus the logarithm of the
second moment $ln(m_{2})$. Three cuts are employed to select the upper branch 
(1), the lower branch (3) and the central region (2).}
\end{figure}
 
\begin{figure}
\label{f16}
\noindent
\caption{Filtered CMD results with selection of events. Impact parameter 
distributions (upper panels) and multiplicity distributions (lower panels) for 
the three cuts made on Fig. 15: left part Cut 1, central part Cut 2 and right 
part Cut 3.}
\end{figure}
 
\begin{figure}
\label{f17}
\noindent
\caption{Filtered CMD results with selection of events. 
Fragment charge distributions (upper panels) and the corresponding scaled 
factorial moments $ln(F_i)$ versus $-ln(\delta s)$ for the three cuts made on 
Fig. 15: left part Cut 1, central part Cut 2 and right part Cut 3. The solid
line on the upper-central panel indicates a power law distribution 
$N(Z) \propto Z^{-\tau}$ with $\tau = 2.2$. In the lower panels, 
solid circles represent the SFM of order $i=2$, open circles $i=3$, open 
squares $i=4$ and open triangles $i=5$.} 
\end{figure}

\begin{figure}
\label{f18}
\noindent
\caption{Experimental results from Ref. [38]. Campi scatter plot. 
The logarithm of the size
of the largest fragment $ln(Z_{max})$ is plotted versus the logarithm of the
second moment $ln(m_{2})$. Three cuts are employed to select the upper branch 
(1), the lower branch (3) and the central region (2).}
\end{figure}
 
\begin{figure}
\label{f19}
\noindent
\caption{Experimental results from Ref. [38]. Scaled 
factorial moments $ln(F_i)$ versus $-ln(\delta s)$ for the three cuts made on 
Fig. 18: left part Cut 1, central part Cut 2 and right part Cut 3. 
Solid circles represent the SFM of order $i=2$, open circles $i=3$, open 
squares $i=4$ and open triangles $i=5$.} 
\end{figure}


\begin{thebibliography}{99}
\bibitem{note} Charge distributions and
the SFM shown in Fig. 17 have been calculated without the heaviest fragment
in each event, apart the charge distribution of Cut 1.
\end{thebibliography}
\end{document}